\documentclass{aastex}
\usepackage{spr-astr-addons}
\usepackage{url}
\usepackage{color}

\RequirePackage{color}

\begin{document}

\title{ 
Resistive and magnetized accretion flows with convection
}

\shorttitle{ Resistive and magnetized accretion flows}
\shortauthors{K. Faghei \& M. Omidvand}

\author{Kazem Faghei and Mobina Omidvand
}
\affil{School of Physics,
Damghan University, Damghan, Iran\\
e-mail: kfaghei@du.ac.ir}

\begin{abstract}
We considered the effects of convection on the radiatively inefficient accretion flows (RIAF) in the presence of resistivity and toroidal magnetic field. We discussed the effects of convection on transports of angular momentum and energy. We established two cases for the resistive and magnetized RIAFs with convection: assuming the convection parameter as a free parameter and using mixing-length theory to calculate convection parameter. A self-similar method was used to solve the integrated equations that govern the behavior of the presented model. The solutions showed that the accretion and rotational velocities decrease by adding the convection parameter, while the sound speed increases. Moreover, by using mixing-length theory to calculate convection parameter, we found that the convection can be important in RIAFs with magnetic field and resistivity.  
\end{abstract}


\section{Introduction}
The existence of radiatively inefficient accretion flows
(RIAFs) have been confirmed in low-luminosity state of X-ray binaries
and nuclei of galaxies (Narayan et al. 1996; Esin et al. 1997; Di Matteo et al. 2003; Yuan et al. 2003).
It was understood that RIAFs are likely to be convectively unstable in
the radial direction due to the inward increase of the entropy of accreting gas (Narayan \& Yi 1994). 
Moreover, hydrodynamical and magetohydrodynamical simulations of low-viscosity RIAFs have confirmed 
these flows are convectively unstable (
e. g. Igumenshchev et al. 1996; Stone et al. 1999; Machida et al. 2001; Hawley \& Balbus 2002; 
McKinney \& Gammie 2002; Igumenshchev et al. 2003).
Self-similar or global solutions for convection-dominated accretion flows
(CDAFs) were presented by several authors (e. g. Narayan et al. 2000; Quataert \& Gruzinov 2000; Abramowicz et al. 2002; 
Lu et al. 2004; Zhang \& Dai 2008). 

Igumenshchev et al. (2003) studied the resistive MHD simulations of RIAFs onto black holes. 
They assumed two cases for the geometry of the injected magnetic field: pure toroidal field and pure poloidal field. 
They found that in the case of pure toroidal magnetic field, the accreting gas forms a nearly axisymmetric, geometrically thick, 
turbulent accretion disc. Moreover, their solutions represented that the flow resembles in many respects CDAFs found in previous
 numerical and analytical investigations of viscous hydrodynamic flows.
Zhang \& Dai (2008) investigated the effect of magnetic field on RIAFs with convection by a semi-analytically method. 
By exploit of $\alpha$-prescription for viscosity and convection, they used two methods
to study of magnetized flows with convection, i.e. they take the
convective coefficient $\alpha_c$ as a free parameter to discuss the effects of convection for simplicity.
They also established the $\alpha_c$-$\alpha$ relation for magnetized flows using the mixing-length theory and
compare this relation with the non-magnetized case. 
They found that the magnetic field makes the $\alpha_c$-$\alpha$ relation be distinct from that of non-magnetized flows. 

Since the importance of toroidal magnetic field and resistivity in accretion flows have been confirmed observationally
(see Faghei 2011 and references therein), Faghei (2011) considered the steady, radially self-similar solutions of  
accretion flows in the presence of the toroidal magnetic field and the resistivity. However, he ignored the effects of convection in his model. 
Generally semi-analytical studies of magnetized CDAFs are related to non-resistive magnetized CDAFs (e. g. Zhang \& Dai 2008)
and the resistive and magnetized CDAF was studied in MHD simulations (e. g. Igumenshchev \& Narayan 2002; 
Hawley \& Balbus 2002; Igumenshchev et al. 2003). 
Thus, it will be interesting to study the effects of resistivity on RIAFs with convection. Here, we adopt the 
presented solutions by Narayan et al. (2000) and Faghei (2011). Similar to Narayan et al. (2000), we will discuss the 
effects of convection on angular momentum and energy equations. The paper is organized as follow. In section 2, the basic 
equations of constructing a model for quasi-spherical magnetized  RIAFs with convection will be defined. In
section 3, a self-similar method for solving equations which govern the behavior of the accreting gas was utilized. The
summary of the model will appear in section 4.

\section{Basic Equations}
Analytical theory of CDAF is based on a self-similar solution of a simplified set of equations describing RIAFs. 
We adopted  the presented solutions by Narayan et al. (2000) and Faghei (2011). By using spherical coordinate 
($r$, $\theta$, $\varphi$) 
centered on a accreting object, let us consider stationary, axisymmetric, quasi-spherical equations describing an accretion flow onto the 
black hole of mass $M$. For the sake of simplicity, the general-relativistic effect has been neglected and the gravitational force on a 
fluid is characterized by Newtonian potential of a point mass, $\psi=-G M / r$. As magnetic fields, we consider only toroidal
fields, $B_\varphi$. 

Under these assumptions, the continuity equation with mass loss is
\begin{equation}
\frac{1}{r^{2}}\frac{d}{d r}(r^{2}\rho v_{r})=\dot{\rho},
\end{equation}
where $\rho$, $v_r$ and $\dot{\rho}$ are the density, the accretion velocity ($v_r < 0$), and the mass-loss per unit volume, respectively.

The radial momentum equation is
\begin{equation}
v_{r}\frac{d v_{r}}{d r} =
  r \left( \Omega^{2}- \Omega^{2}_K\right)-\frac{1}{\rho}\frac{d }{d r}(\rho c^2_s)-\frac{c^2_A}{r}-\frac{1}
     {2 \rho}\frac{d }{d r}(\rho c^2_A),
\end{equation}
where $c_s$ is sound speed, which is defined as $c_s^2\equiv p_{gas} / \rho$, with being $p_{gas}$ as the gas pressure,
$\Omega$ is the angular velocity, $\Omega_K [=\left( G M / r^3\right)^{1/2}]$ is the Keplerian angular velocity, 
and $c_A$ is the alfven speed, which is defined as $c_A^2\equiv B_\varphi^2 / 4 \pi \rho = 2 p_{mag} / \rho$, 
with being $p_{mag}$ as the magnetic pressure.
The ram-pressure term $v_r d v_r / d r$ and last two terms due to the magnetic field in this equation were ignored
 in the self-similar CDAF model of Narayan et al. (2000), while we include them here in order to consider their effects.

The angular momentum equations can be written in the form of the balance of advection and diffusion transport terms (Narayan et al. 2000),
\begin{eqnarray}
\nonumber \rho v_{r}\frac{d}{d r}(r^{2}\Omega)=
  \frac{1}{r^{2}}\frac{d}{d r}\left[\nu \rho r^{4}\frac{d \Omega}{\partial r}\right]+ ~~~~~~~~~~\\ 
\frac{1}{r^{2}}\frac{d}{d r}\left[\nu_{c} \rho r^{(5+3 g)/2}\frac{d }{d r} \left( \Omega r^{3(1-g)/2} \right)\right],
\end{eqnarray}
where the two terms of right hand side represent the angular momentum transport by viscosity and convection. 
Here, $\nu$ is the kinematic viscosity coefficient, $\nu_c$ is the convective diffusion coefficient, and
$g$ is the parameter to determine  the condition of convective angular momentum transport. 
When $g=1$, the flux of angular momentum due to convection is
\begin{equation}
 \dot{J}_c=-\nu_{c}\,\rho\,r^4\,\frac{d \Omega}{d r}.
\end{equation}
The above equation implies that the convective angular momentum flux is oriented down the angular velocity gradient.
For a quasi-Keplerian angular velocity, $\Omega \propto r^{-3/2}$,
angular momentum is transported outward.
When $g=-1/3$, the convective angular momentum flux can be written as
\begin{equation}
 \dot{J}_c=-\nu_{c}\,\rho\,r^2\,\frac{d \left( \Omega r^2 \right)}{d r}.
\end{equation}
 This equation represents that the convective angular momentum flux is
oriented down the specific angular momentum gradient.
For a quasi-Keplerian angular velocity, $\Omega \propto r^{-3/2}$, 
angular momentum is transported inward.
Generally, convection transports angular momentum inward (or outward) for $g<0$ (or $>0$), 
and the specific case $g=0$ corresponds to zero angular momentum transport (Narayan et al. 2000).

In this paper, we assume 
the kinematic coefficient of viscosity and the magnetic diffusivity due to turbulence 
in the accretion flow. So, we use these parameters in analogy to the 
$\alpha$-prescription of Shakura \& Sunyaev (1973) for the turbulent,
\begin{equation}
   \nu = P_m \eta = \alpha \frac{c_s^2}{\Omega_{K}},
\end{equation}
where $P_m$ is the magnetic Prandtl number of the turbulence, which assumed to be a  constant less than unity, 
$\eta$ is the magnetic diffusivity, 
and $\alpha$ is a free parameter less than unity.
For the convective diffusion coefficient, $\nu_c$, we adopt the assumptions of 
Narayan et al. (2000)  and Lu et al. (2004)
that all transport phenomena due to convection have the same diffusion coefficient, which is defined as
\begin{equation}
 \nu_c=\left(\frac{L_M^2}{4}\right)\sqrt{-N_{eff}^2},
\end{equation}
where $L_M$ is the characteristic mixing length and $N_{eff}$ is the effective frequency of convective blobs. 
The characteristic mixing length $L_M$
in terms of the pressure scale height, $H_p$, can be written as
\begin{equation}
 L_M=2^{-1/4} l_M H_p, ~~~~ H_p=-\frac{d r}{d \ln p_{gas}},
\end{equation}
where $l_M$ is the dimensionless mixing-length parameter and its amount is estimated to be equal to $\sqrt{2}$ in ADAFs 
(Narayan et al. 2000; Lu et al. 2004). the effective frequency of convective blobs, $N_{eff}$, is given by
\begin{equation}
 N_{eff}^2=N^2+\kappa^2,
\end{equation}
where $N$ is Brunt-V\"ais\"al\"a frequency, which is defined as
\begin{equation}
N^2=-{1\over\rho}{dp_{gas}\over dr}{d\over dr}\ln\left({p_{gas}^{1/\gamma}\over\rho}
\right), 
\end{equation}
and $\kappa$ is epicyclic frequency, which is defined as 
\begin{equation}
\kappa^2=2 \Omega^2 \frac{d \ln (\Omega r^2)}{d \ln r}.
\end{equation}
For a non-Keplerian flows $\kappa \neq \Omega$, while for a quasi-Keplerian ($\Omega \propto r^{-3/2}$),
$\kappa = \Omega$ 
(Narayan et al. 2000; Lu et al. 2004). Convection appears in flows with $N_{eff}^2 < 0$.
We also write the convective diffusion coefficient in the form similar to usual viscosity of Shakura \& Sunyaev (1973),
\begin{equation}
\nu_c=\alpha_c \frac{c_s^2}{\Omega_K}
\end{equation}
where $\alpha_c$ is a dimensionless coefficient that describes the strength of convective diffusion. 
The $\alpha_c$ coefficient can be obtained by equations (8) and (13)
\begin{equation}
\alpha_c =\frac{\Omega_K}{c_s^2} \left(\frac{L_M^2}{4}\right)\sqrt{-N_{eff}^2}.
\end{equation}

The energy equation is 
\begin{eqnarray}
 \nonumber
\rho v_r T \frac{d s}{d r}\equiv \rho v_r \left[ \frac{1}{\gamma-1}
\frac{d c_s^2}{d r} - \frac{c_s^2}{\rho} \frac{d \rho}{d r}  \right]= \\ 
 Q_{diss}+Q_{conv}-Q_{rad},
\end{eqnarray}
where $T$ is the temperature, $s$ is the specific entropy, $\gamma$ is the ratio of specific heats,
$Q_{diss}$ is dissipative heating rate, $Q_{rad}$ is the radiative cooling rate, and 
$Q_{conv}=-\mathbf \nabla\cdot \mathbf F_{conv}$, with being
$F_{conv}[=-\rho\nu_c T d s/d r]$ as the outward energy flux due to convection.
For the right hand side of the energy equation, we can write
\begin{eqnarray}
 Q_{adv}=f Q_{diss}-\frac{1}{r^2} \frac{d}{dr}\left(r^2 F_{conv} \right),
\end{eqnarray}
where $Q_{adv}$ is the advective transport of energy, and
$f [= 1 - Q_{rad}/Q_{diss}]$ is the advection parameter. The parameter $f$ measures the
degree to which the flow is advection-dominated (Narayan \& Yi 1994).
The dissipative heating rate can be written as
\begin{equation}
Q_{diss}=(\nu+g \nu_c)\rho r^2 \left(\frac{\partial \Omega}{\partial  r}\right)^2+\frac{\eta}{4\pi} {\bf J}^2,
\end{equation}
where the right-hand side terms are  heating rate due to viscosity, convection, and resistivity, respectively.
In above equation, ${\mathbf J} [=\nabla\times {\mathbf B}]$ is the current density, 
with being ${\mathbf B}$ as the magnetic field.

Finally, the induction equation with creation/escape of magnetic field can be written as
\begin{equation}
\frac{1}{r}\frac{d
}{d r}\left[r v_{r} B_{\varphi}-\eta\frac{d
}{d r}(r B_{\varphi})\right]=\dot{B}_{\varphi}.
\end{equation}
where $B_{\varphi}$ is the toroidal component of
magnetic field and $\dot{B}_{\varphi}$ is the field escaping/creating rate due to 
a magnetic instability or dynamo effect. This induction equation is rewritten as
\begin{eqnarray}
\nonumber\dot{B}_{\varphi}= ~~~~~~~~~~~~~~~~~~~~~~~~~~~~~~~~~~~~~~~~~~~~~~~~~~~~~~~~~\\
\frac{1}{r}\frac{d
}{d r}\left[\sqrt{4\pi\rho c^2_A}\left(r v_{r} -\frac{\alpha}{4 \chi P_m }\frac{1}{ r \rho\Omega_K}\frac{d
}{d r}(r^2 \rho c^2_A )\right)\right],
\end{eqnarray}
where $\chi$ is the ratio of the magnetic pressure to the gas pressure, which is defined by
\begin{equation}
\chi=\frac{p_{mag}}{p_{gas}}=\frac{1}{2} \left(\frac{c_A}{c_s}\right)^2.
\end{equation}

\section{Self-Similar Solutions}
We seek self-similar solutions in the following form (e.g.
Narayan \& Yi 1994; Akizuki \& Fukue 2006)
\begin{equation}
v_r(r)=-c_1 \alpha \sqrt{\frac{G M_*}{r}}
\end{equation}
\begin{equation}
\Omega(r)=c_2\sqrt{\frac{G M_*}{r^3}}
\end{equation}
\begin{equation}
c^2_s(r)=c_3\frac{G M_*}{r}
\end{equation}
\begin{equation}
c^2_A(r)=\frac{B^2_{\varphi}}{4\pi\rho}=2 \chi c_3\frac{G M_*}{r}
\end{equation}
where $c_1$, $c_2$, and $c_3$ are dimensionless constant to be determined. We use a power-law relation for density
\begin{equation}
\rho(r)=\rho_0 r^\lambda,
\end{equation}
where $\rho_0$ and $\lambda$ are constant. Using equations (20)-(24), the mass-loss 
rate and the magnetic field escaping/creating rate can be written as

\begin{equation}
\dot{\rho}(r)=\dot{\rho}_0 r^{\lambda-3/2},
\end{equation} 

\begin{equation}
\dot{B}_{\varphi}(r)=\dot{B}_0 r^{\frac{\lambda-4}{2}},
\end{equation}
where $\dot{\rho}_0$ and $\dot{B}_0$ are constant. Since we have not applied the effects
of wind in the momentum and energy equations, we will 
assume a no wind case, $\dot\rho=0$ and $\lambda=-3/2$. 
In this case, $\dot{B}_{\varphi}\propto r^{-11/4}$, 
which implies that
creation/escape of magnetic field increases with approaching to central object. 
This property is qualitatively consistent with previous studies
of accretion flows (Machida et al. 2006; Oda et al. 2007; Faghei \& Mollatayefeh 2012).

Using the self-similar solutions in the continuity, radial momentum, angular momentum,  convection parameter,
energy, and induction equations [(1)-(3), (13), (14), and (18)], we can obtain the following relations:

\begin{equation}
\dot{\rho}_0=-\left(\lambda+\frac{3}{2}\right) \alpha \rho_0 c_1 \sqrt{G M_*},   
\end{equation}
\begin{equation}
-\frac{1}{2}c^2_1 \alpha^2 + 1 - c^2_2 + c_3 \left[\lambda-1+\chi (1+\lambda)\right]=0    ,
\end{equation}
\begin{equation}
\alpha c_1 = 3 (\alpha+ g \alpha_c) (\lambda+2) c_3   ,
\end{equation}
\begin{eqnarray}
\nonumber \alpha c_1 \left[\frac{1}{\gamma-1}+\lambda\right]=~~~~~~~~~~~~~~~~~~~~~~~~~~\\
\nonumber \frac{9}{4}\alpha f \left[
(1+\frac{\alpha_c}{\alpha}g) c_2^2+\frac{2\chi}{9 P_m} c_3 (1+\lambda)^2
\right]\\
-\alpha_c c_3 (\lambda+\frac{1}{2}) \left[\frac{1}{\gamma-1}+\lambda\right]
\end{eqnarray} 
\begin{equation}
\alpha_c= \frac{l_M^2}{4\sqrt{2} c_3 (\lambda-1)^2} \sqrt{\frac{c_3 (\lambda-1)}{\gamma} [\lambda (1-\gamma)-1]-c_2^2} ,  
\end{equation} 
\begin{equation}
\dot{B}_0=-\frac{\alpha \lambda}{2} G M_* \sqrt{2\pi\rho_0\chi c_3}\left[
2 c_1  + \frac{c_3}{P_m}  (1+\lambda) \right].  
\end{equation} 
Above equations express for $\lambda=-3/2$, there is no mass loss, while for $\lambda > -3/2$ mass loss (wind) exists.

\section{Results}
Here, similar to Zhang \& Dai (2008), we will study the presence of convection in two cases: $\alpha_c$ as a free parameter
and $\alpha_c$ as a variable.

%#####################################
\input{epsf}
\begin{figure*}[!ht]
\begin{center}
\centerline
{ 
{\epsfxsize=5.5cm\epsffile{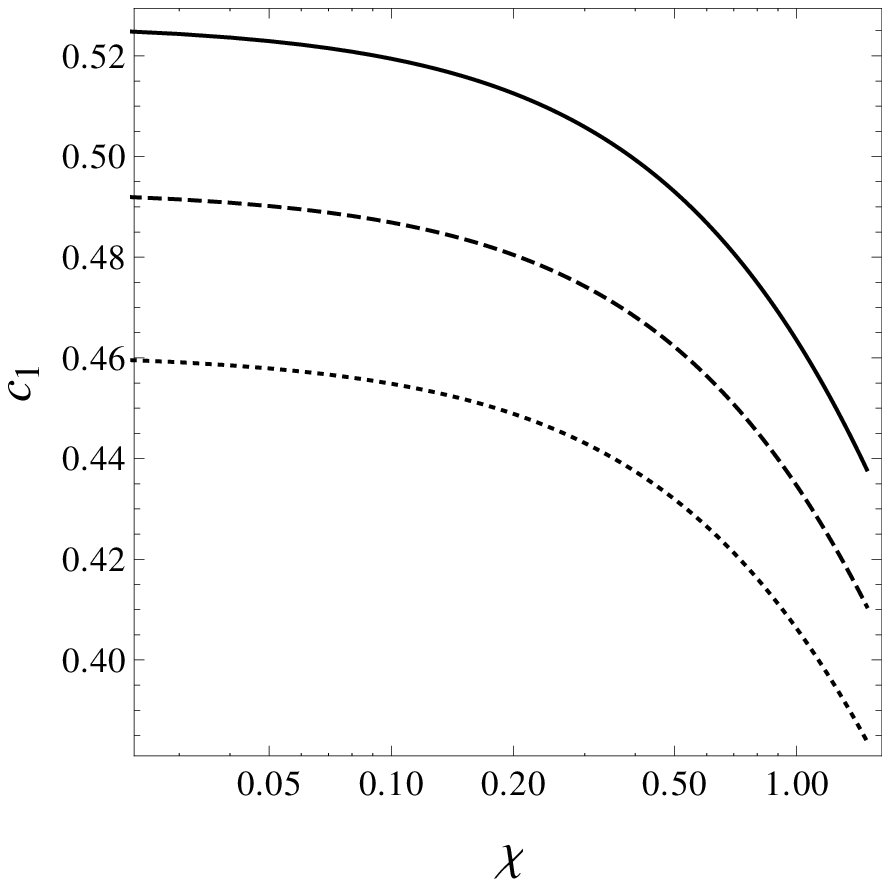} } ~~{\epsfxsize=5.5cm\epsffile{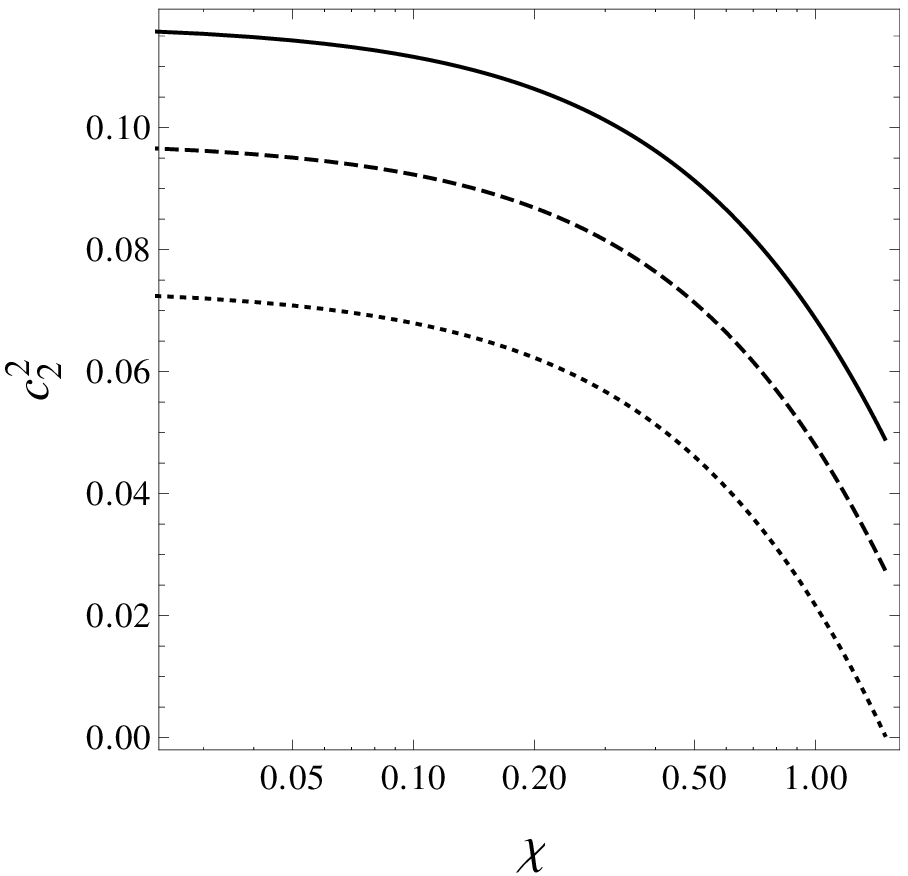}} ~~{\epsfxsize=5.5cm\epsffile{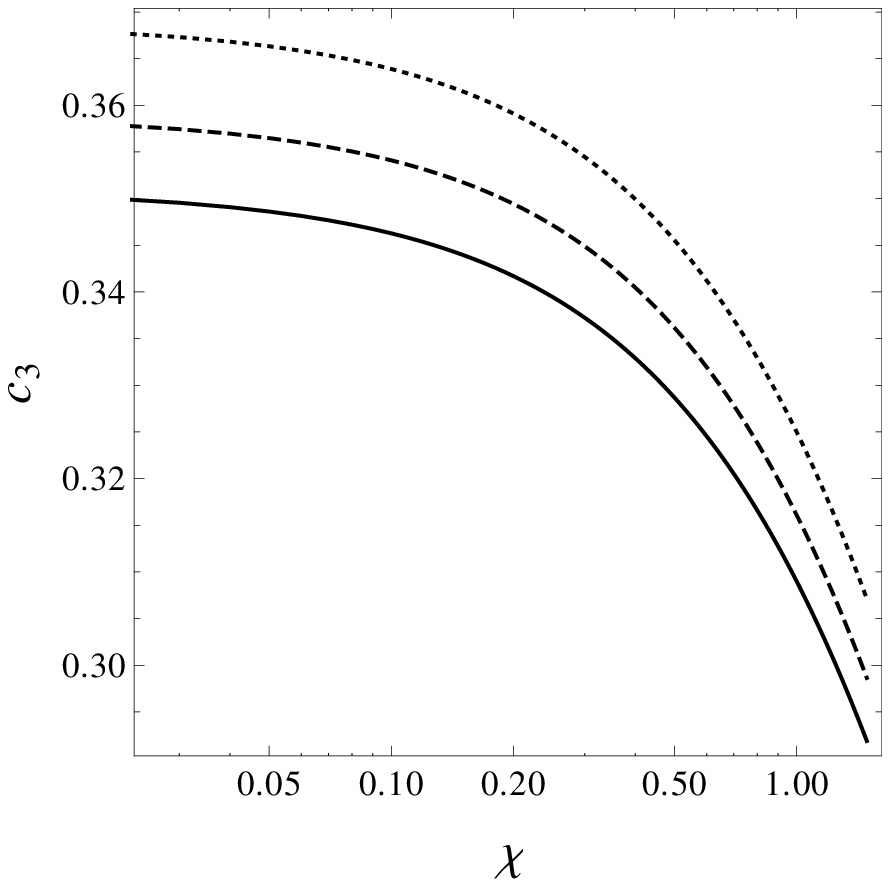}}
} 
\end{center}
\begin{center}
\caption{Physical variables as functions of $\chi$ for several values of convective viscosity. The input parameters are set to
$\alpha=0.2$, $\gamma=1.5$, $P_m=1/2$, $f=1$, $l=\sqrt{2}$, $g=-1/3$, and $\lambda=-3/2$. The solid, dashed, and dotted lines represent 
$\alpha_c=0$, $0.05$, and $0.1$, respectively. }
\end{center}
\end{figure*}
%#####################################

%#####################################
\input{epsf}
\begin{figure*}[!ht]
\begin{center}
\centerline
{ 
{\epsfxsize=5.5cm\epsffile{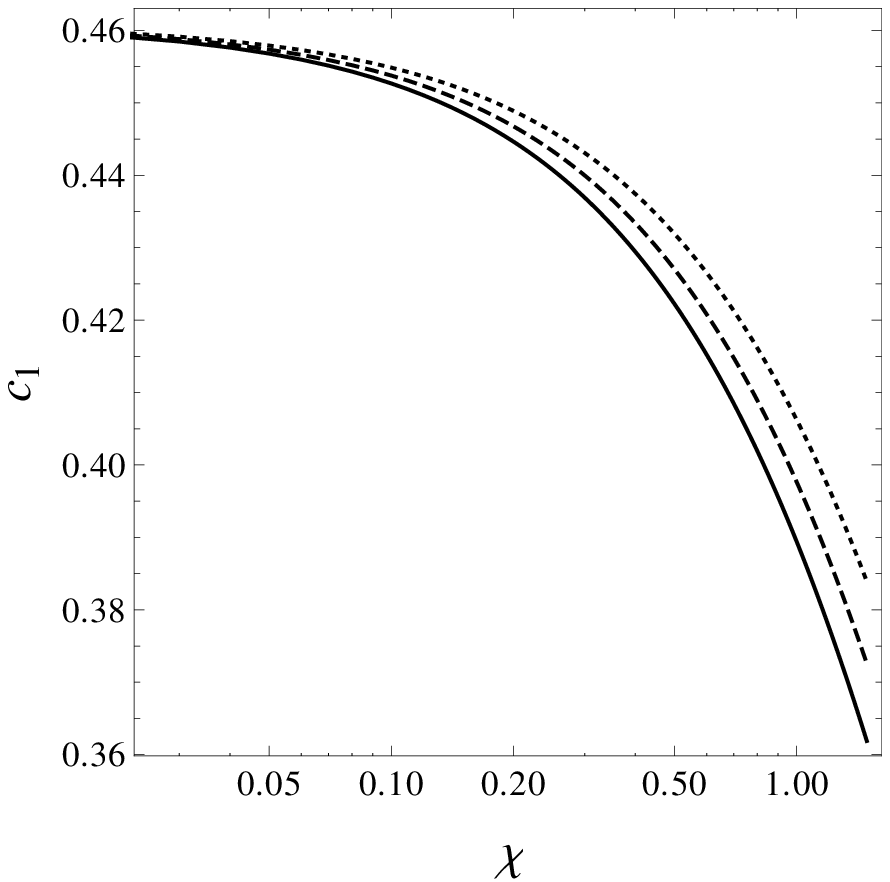} } ~~{\epsfxsize=5.5cm\epsffile{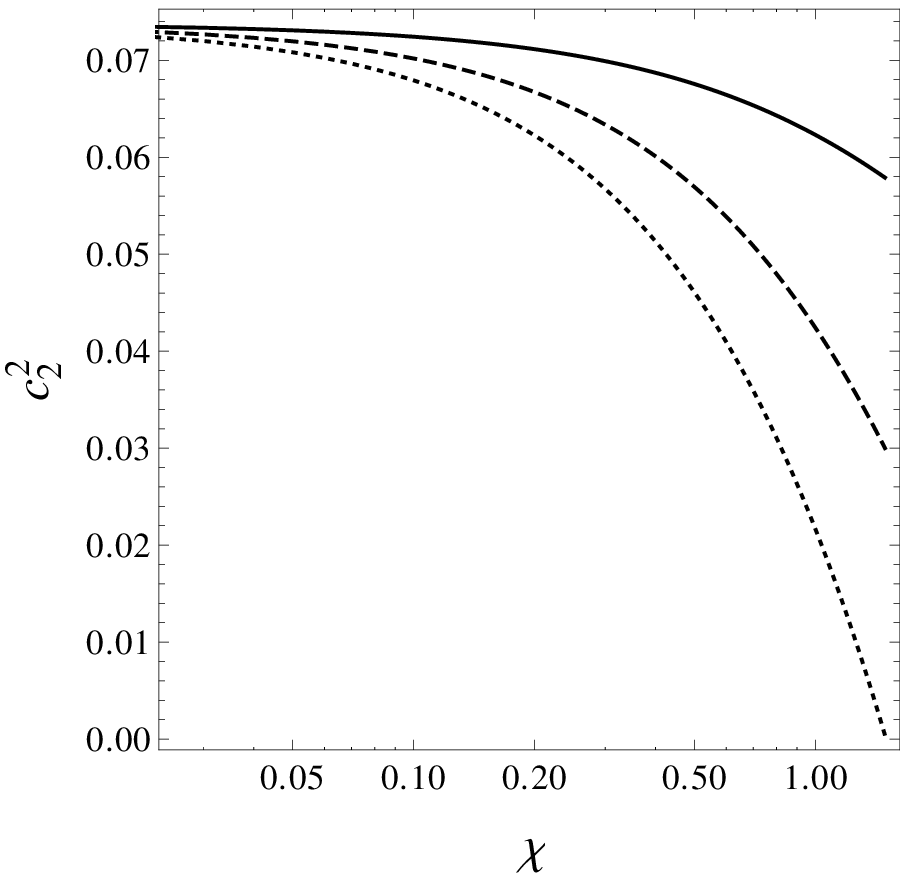}} ~~{\epsfxsize=5.5cm\epsffile{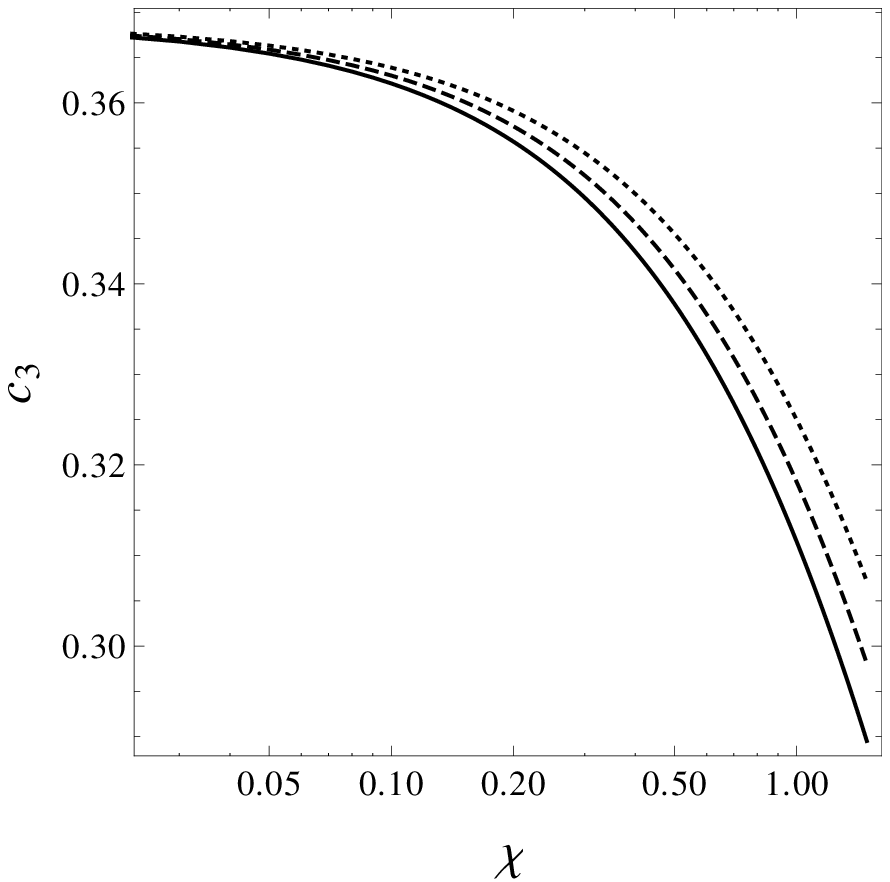}}
} 
\end{center}
\begin{center}
\caption{Same as Figure 1, but $\alpha_c=0.1$, and the solid, dashed, and dotted lines represent 
$P_m=\infty$, $1.0$, and $0.5$, respectively.
}
\end{center}
\end{figure*}
%#####################################

%#####################################
\input{epsf}
\begin{figure*} [ht]
\begin{center}
\centerline
{ 
{\epsfxsize=6.7cm\epsffile{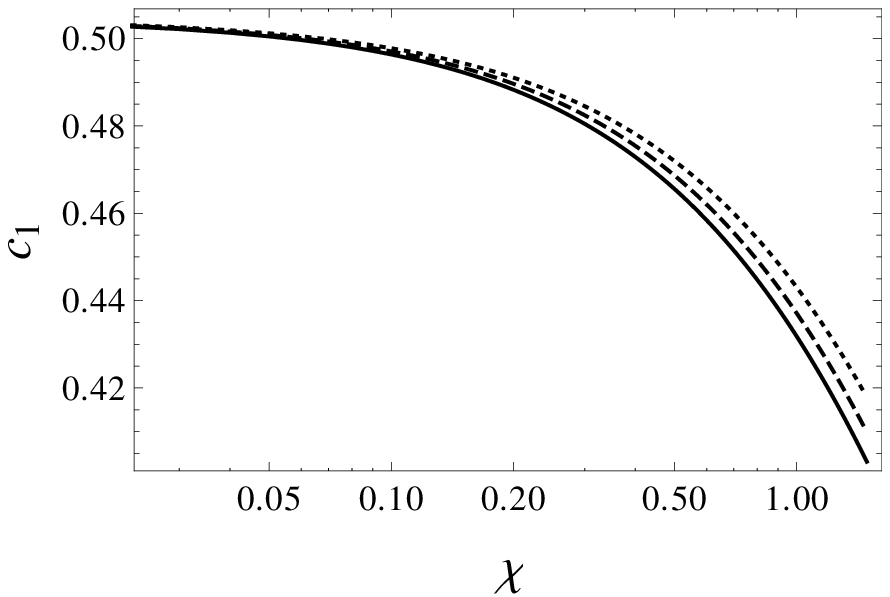} }~~~~~~~~{\epsfxsize=6.7cm\epsffile{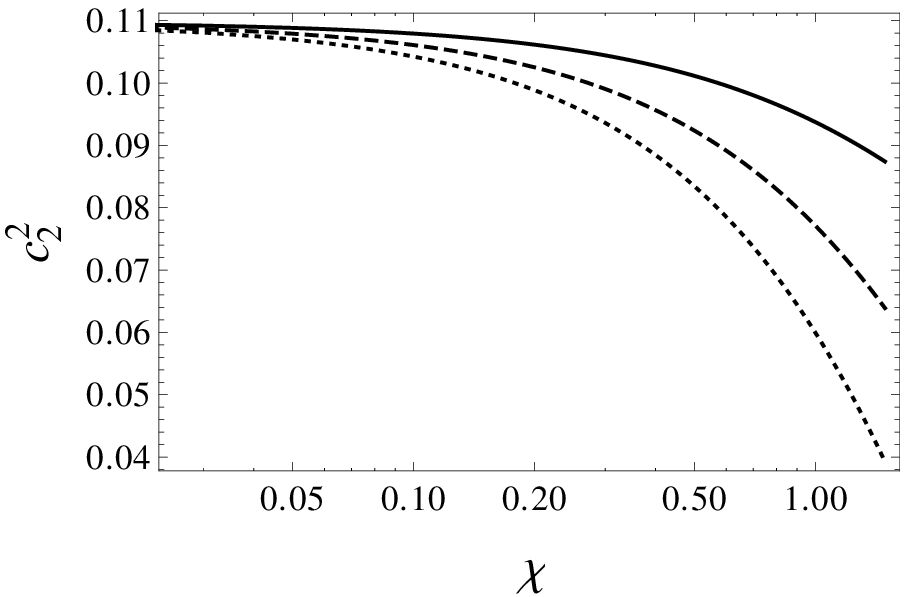}  }
} 
\centerline
{ 
{\epsfxsize=6.7cm\epsffile{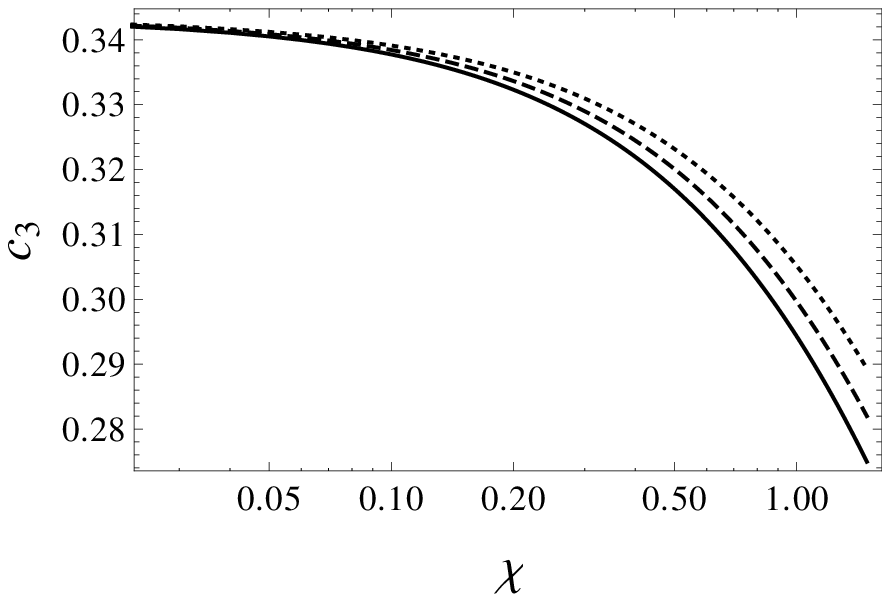} }~~~~~~~~{\epsfxsize=6.7cm\epsffile{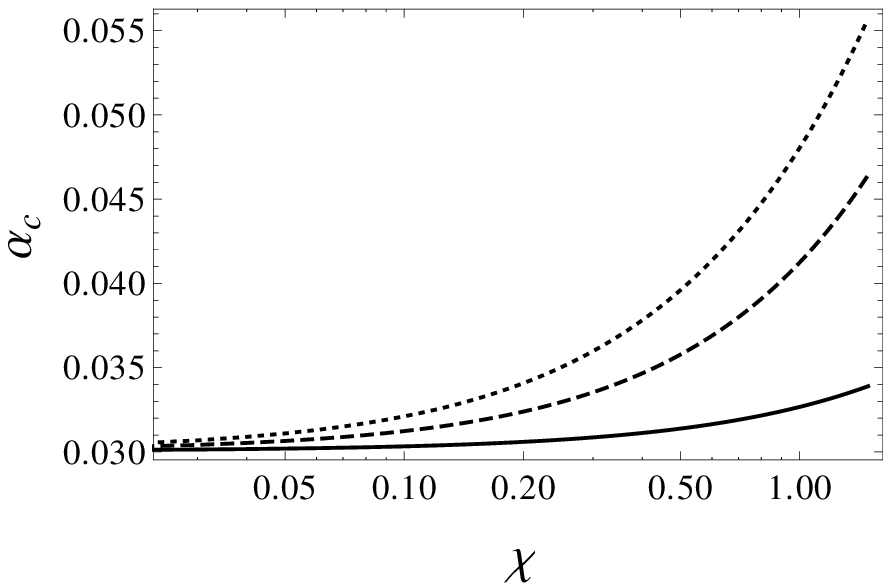}  }
} 
\end{center}
\begin{center}
\caption{Physical variables as functions of $\chi$ for several values of magnetic Prandtl number. 
The input parameters are set to
$\alpha=0.5$, $\gamma=1.5$, $f=1$, $l=\sqrt{2}$, $g=-1/3$, and $\lambda=-3/2$. The solid, dashed, and dotted lines represent 
$P_m=\infty$, $1.0$, and $0.5$, respectively.}
\end{center}
\end{figure*}
%#####################################

\subsection{\textit{Case 1: $\alpha_c$ as a free parameter}}
In this case, we take the
convective coefficient $\alpha_c$ as a free parameter to discuss the effects of convection for simplicity.
Examples of such solutions are presented in Figures 1 and 2.

In Figure 1, the self-similar coefficients $c_1$, $c_2$, and $c_3$ are shown as
functions of the parameter $\chi$.
By adding the parameter $\chi$ which indicates the role of
magnetic filed on the dynamics of accretion discs, we see the coefficients of radial and rotational velocities and sound speed decrease.
This properties are qualitatively consistent with results of Faghei (2011). In Figure 1, we also studied the effect of 
convection parameter $\alpha_c$ on the physical variables. The value of $\alpha_c$ measures the strength
of convective viscosity and a larger $\alpha_c$ denotes a stronger turbulence due to convection.
Figure 1 implies that for non-zero $\alpha_c$, the radial infall velocity is lower
than the standard ADAF solution and for larger $\alpha_c$  this reduction of radial infall velocity is more evident. 
It can be due to decrease of efficiency of angular momentum transport by adding the convection parameter $\alpha_c$ (see equation 29). 
The profiles of angular velocity show that it decreases with the magnitude of $\alpha_c$, while the sound speed increases.
These properties are in accord with results of Zhang \& Dai (2008).

In Figure 2, the physical variables are shown as functions of parameter $\chi$ for several values of magnetic Prandtl number.
Since inverse of magnetic Prandtl number is proportional to magnetic diffusivity, $P_m \propto \eta^{-1}$. Thus, reduce of magnetic Prandtl number 
denotes to increase of resistivity of the fluid. The solutions in Figure 2 imply that the accretion velocity and the sound speed both increase
with the magnitude of resistivity, while the rotational velocity decreases. These properties qualitatively confirm the results of Faghei (2011).

\subsection{\textit{Case 2: $\alpha_c$ as a variable}}
Here, we calculate the dimensionless coefficient $\alpha_c$ by using the mixing-length theory. Because we used a steady self-similar
method to derive $\alpha_c$, it becomes a constant throughout of the accreting gas. However, it is a function of position 
and time (e. g. Lu et al. 2004).
The amount of convection parameter $\alpha_c$ is calculated by equation (31). Using this equation and equation (28)-(30), we can
obtain the behavior of physical quantities in the presence of convection. Such solutions are shown in Figure 3.
 
In Figure 3, the coefficients $c_1$, $c_2$, $c_3$, and convection parameter $\alpha_c$ are shown as functions of the degree of magnetic pressure.
Similar to case 1, the accretion and rotational velocities, and sound speed decrease by adding the parameter $\chi$.
While, the convection parameter $\alpha_c$ increases for stronger toroidal magnetic field. This property is qualitatively consistent with
result of Zhang \& Dai (2008). In Figure 3, the physical variables are also studied for several values of magnetic Prandtl number. 
The profiles of convection parameter $\alpha_c$ imply that it increases by adding the magnetic diffusivity. As for non-zero magnetic diffusivity,
$\alpha_c$ is larger than the standard CDAF solution and for larger magnetic diffusivity this increase of convection parameter $\alpha_c$
is more evident. 
 
\section{Summary and Discussion}
The observational features of
low-luminosity state of X-ray binaries and nuclei of galaxies can be successfully explained by the 
models of radiatively inefficient accretion flow (RIAF). The importance of convection in RIAFs was realized
by semi-analytical and direct numerical simulation (e. g. Narayan et al. 2000; Igumenshchev et al. 2003).  

In this research, we considered the effects of convection on the presented model of Faghei (2011). Similar to Narayan et al. (2000), 
we assumed the convection affects on transports of angular momentum and energy. Using a radially self-similar approach, we studied 
the effects of convection on the model for several values of magnetic field and resistivity. The solutions showed that the accretion 
and rotational velocities, and sound speed decrease for stronger magnetic filed. Moreover, we found that the accretion velocity 
and sound speed increase with the magnitude of the resistivity, while the rotational velocity decreased. These properties are 
qualitatively consistent with results of Faghei (2011). We studied the effects of convection on a resistive and magnetized RIAF in two cases:
assuming the convection parameter as a free parameter and using mixing length theory to calculate the convection parameter.
In the first case, we found that by adding the convection parameter, the radial and rotational velocities decrease and
the sound speed increases. In the second case, we found that the convection parameter increases by adding the magnetic filed and
resistivity. These properties are in many aspects in accord with results of Zhang \& Dai (2008).    

The present model have some limitations that can be modified in the future works.
For example, the latitudinal dependence of physical variables have been ignored in this paper. 
While, two-dimensional and three-dimensional MHD simulations of RIAFs show that 
the disc geometry strongly depends
on magnetic field configuration (e. g. Igumenshchev et al. 2003). Thus, the study of present model in two/three dimensions 
can be an interesting subject for future research. 
Moreover, it has been understood the magnetic field can change the criterion
for convective instability (e. g. Balbus \& Hawley 2002). 
While, we igonred the effects of magnetic field on the instability criterion.
Thus, the presented criterion in this paper can be modified in
the future research.

\section*{Acknowledgements}
I wish to thank the anonymous referee for very useful comments
that helped us to improve the initial version of the paper.

\end{document}